\documentclass[showpacs,floatfix,amsmath,amssymb,nofootinbib,superscriptaddress]{ws-ijmpa}
\usepackage{graphicx}
\usepackage{epstopdf}
\usepackage{bbm}
\usepackage{xcolor}
\usepackage{amsmath}
\usepackage{slashed}
\usepackage{float}
\usepackage{subfigure}
\usepackage{hyperref}
\begin{document}

\title{FCNC contributions to the $Z\gamma V^\ast$ vertex, polarizations and bounds on $Z\overline{t}q$ couplings}
\author{A.I. Hern\'andez-Ju\'arez}
\address{Departamento de F\'isica, FES-Cuautitl\'an, Universidad Nacional Aut\'onoma de M\'exico\\ C.P. 54770, Estado de M\'exico, M\'exico. \\Facultad de Ciencias F\'isico Matem\'aticas, Benem\'erita Universidad Aut\'onoma de Puebla\\
 Apartado Postal 1152, Puebla, Pue., M\'exico.\\alan.hernandez@cuautitlan.unam.mx}
\author{R. Gait\'an}
\address{Departamento de F\'isica, FES-Cuautitl\'an, Universidad Nacional Aut\'onoma de M\'exico\\ C.P. 54770, Estado de M\'exico, M\'exico.}
\author{G. Tavares-Velasco}
\address{Facultad de Ciencias F\'isico Matem\'aticas, Benem\'erita Universidad Aut\'onoma de Puebla\\ Apartado Postal 1152, Puebla, Pue., M\'exico.}

\maketitle

\begin{history}
\received{Day Month Year}
\revised{Day Month Year}
\accepted{Day Month Year}
\published{Day Month Year}
\end{history}

\begin{abstract}
We study the flavor-changing neutral current contributions to the trilinear neutral gauge boson couplings $Z\gamma V^\ast$ ($V=Z$, $\gamma$) at the one-loop level. Only the $CP$-conserving form factor $h_3^Z$ is induced, with the relevant contributions arising from top quark couplings. To our numerical analysis, constraints on the $Z\overline{t}q$ ($q=c$, $u$) couplings are obtained from current LHC data, leading to bounds of $|g_{\text{V}}^{tu}|\text{,}|g_{\text{A}}^{tu}|\leqslant 0.007$ and  $|g_{\text{V}}^{tc}|\text{,}|g_{\text{A}}^{tc}|\leqslant 0.0095$ for the vector and axial couplings. The new contributions to the $h_3^Z$ form factor are of order $10^{-7}$. Furthermore, we analyze the polarized partial decay widths of the $V^\ast\rightarrow Z\gamma$ process, and a new type of left-right asymmetry is also discussed. We find that in the presence of new physics, the polarized $\Gamma(V^\ast\rightarrow Z\gamma)$ exhibits significant deviations from the SM prediction, resulting in a non-zero asymmetry. 
 \end{abstract}



\section{Introduction}
The CMS and  ATLAS collaborations have recently utilized data collected from LHC at $\sqrt{s}=$ 13 TeV to set the latest limits on trilinear neutral gauge boson couplings (TNGBCs) $Z\gamma V$ ($V=\gamma, Z$) \cite{CMS:2020gtj,ATLAS:2018nci}. These constraints closely align with the predictions of the Standard Model (SM)  \cite{Gounaris:2000tb,Choudhury:2000bw} and represent one-order-of-magnitude improvement compared to previous LHC analyses \cite{CMS:2017dzg,ATLAS:2016qjc}. With the forthcoming luminosity upgrade for Run 3 at the LHC, there is an opportunity for increased sensitivity to TNGBCs, making them potentially measurable soon. Therefore, it will be essential to study the implications of TNGBCs on the LHC and future collider experiments to analyze upcoming results.

Due to Bose statistics and angular momentum conservation, non-vanishing TNGBCs require at least one off-shell gauge boson. These couplings can be induced at the one-loop level or higher order through dimension-six and dimension-eight operators, respectively \cite{Gounaris:2000tb,Larios:2000ni,Gounaris:2000dn,Degrande:2013kka}. TNGBCs of the $Z\gamma V$ ($V=\gamma, Z$) type can be parametrized by two $CP$-conserving and two $CP$-violating form factors. The former can be generated at the one-loop level within the SM, whereas those associated with $CP$ violation may be non-zero in new physics models \cite{Gounaris:2000tb,Hernandez-Juarez:2021mhi}. 
The first experimental bounds on TNGBCs were obtained from data collected at LEP \cite{Acciarri:1998iw,Abbiendi:2000cu,Abdallah:2007ae} and Tevatron \cite{Abe:1994fx,Abachi:1997xe,Aaltonen:2011zc} colliders. The phenomenology of these couplings has been analyzed in earlier studies \cite{Gounaris:1999kf,Baur:1992cd,Alcaraz:2001nv,Walsh:2002gm,Atag:2004cn,MoortgatPick:2005cw,Gounaris:2005pq,GutierrezRodriguez:2008tb} and remains an area of active research due to a renewed interest in the topic \cite{Ananthanarayan:2011fr,Senol:2013ym,Ananthanarayan:2014sea,Chiesa:2018lcs,Ellis:2019zex,Lombardi:2021wug}.
Additionally, the effects of TNGBCs on future colliders have been explored in recent studies  \cite{Rahaman:2017qql,Rahaman:2016pqj,Behera:2018ryv,Rahaman:2018ujg,Rahaman:2020jll,Yilmaz:2021qnv,Spor:2022zob,Yang:2021kyy}.
On the theoretical side, contributions from several SM extensions to TNGBCs have been examined, including the minimal supersymmetric standard model (MSSM)\cite{Gounaris:2000tb,Choudhury:2000bw,Gounaris:2005pq}, the $CP$-violating two-Higgs doublet model (2HDM) \cite{Corbett:2017ecn,Belusca-Maito:2017iob,Grzadkowski:2016lpv}, models with axial and vector fermion couplings \cite{Corbett:2017ecn}, as well as those with an extended scalar sector \cite{Dutta:2009nf,Moyotl:2015bia},  models incorporating $CP$ violation \cite{Biekotter:2021int}, and through an effective Lagrangian approach \cite{Larios:2000ni}. Recently, TNGBCs involving the $ZZV^\ast$ ($V=\gamma$, $Z$) vertex were explored in models that allow flavor-changing neutral currents (FCNCs) \cite{Hernandez-Juarez:2021mhi}. In this context, the $CP$-violating form factors can arise when complex FCNC couplings are considered. 

 On the other hand, the polarizations of neutral gauge bosons, along with the absorptive part of anomalous couplings, can be useful in the search for new physics effects \cite{Maina:2020rgd,Maina:2021xpe,Hernandez-Juarez:2024zpk}. Particularly noteworthy are the newly defined left-right asymmetries in polarized amplitudes of gauge bosons. These asymmetries require complex and $CP$-violating form factors to yield non-zero results  \cite{Hernandez-Juarez:2023dor}. In the SM, the form factors that parametrize the TNGBCs are complex \cite{Gounaris:2000tb}, indicating that these asymmetries may also manifest within the context of TNGBCs. Previous studies have examined polarization effects on TNGBCs \cite{Rizzo:1999xj,Atag:2003wm,Ananthanarayan:2004eb,Ananthanarayan:2011fr,Ananthanarayan:2014sea,Rahaman:2017qql,Subba:2023jia,Atag:2004cn,Rahaman:2016pqj,Rahaman:2018ujg,Jahedi:2022duc,Jahedi:2023myu}, where some angular asymmetries have been proposed as sensitive observables for detecting new physics  \cite{Rizzo:1999xj,Ananthanarayan:2004eb,Ananthanarayan:2014sea,Rahaman:2016pqj,Rahaman:2017qql,Rahaman:2018ujg}. Moreover, the study of polarized neutral gauge bosons at the LHC is particularly interesting. For instance, the ATLAS collaboration has reported measurements of longitudinally polarized pairs of $Z$ bosons \cite{ATLAS:2023zrv}, and the polarization fractions of the $Z$ boson have also been measured by the ATLAS, CMS, and LHCb collaborations \cite{ATLAS:2016rnf,ATLAS:2023lsr,CMS:2015cyj,LHCb:2022tbc}. Additionally, the LHCb collaboration has measured photon polarization in the decays $B^0_s\rightarrow \phi\gamma$ and $\Lambda^0_b\rightarrow\Lambda\gamma$ \cite{LHCb:2019vks,LHCb:2021byf}, with indirect measurements in $B^0$ decays conducted by the Belle and BaBar experiments  \cite{Belle:2006pxp,BaBar:2008okc}. The polarizations of the $W^\pm$ gauge boson have also been studied at the LHC \cite{CMS:2011kaj,ATLAS:2012au,ATLAS:2016fbc,CMS:2016asd,ATLAS:2019bsc,CMS:2020ezf,CMS:2020etf,ATLAS:2022rms}.  Finally, event generators such as \texttt{MadGraph5\_aMC@NLO} \cite{BuarqueFranzosi:2019boy} and \texttt{SHERPA} \cite{Hoppe:2023uux} have also included the capability of generate polarized amplitudes.

In this work, we investigate whether, analogous to the  $ZZV^\ast$ ($V=Z$, $\gamma$) vertex \cite{Hernandez-Juarez:2021mhi}, $CP$-violating contributions in the $Z\gamma V^\ast$ ($V=Z$, $\gamma$) scenario can arise from a general model with complex FCNC couplings mediated by the $Z$ boson. Furthermore, following the approach outlined in Ref. \refcite{Hernandez-Juarez:2023dor}, we calculate the polarized amplitudes of the $V^\ast\rightarrow Z\gamma$ process to uncover new asymmetries associated with transversally polarized gauge bosons. These observables may be sensitive to physics beyond the SM. The rest of the presentation is organized as follows. In Sec. \ref{teo}, we present a review of the TNGBCs of  type
$Z\gamma V^\ast$ ($V=Z$, $\gamma$) and introduce the basics of FCNCs couplings. We discuss in Sec. \ref{analyticSec} the main steps of our calculations for the FCNC contributions to the TNGBCs. We also compute the unpolarized and polarized partial widths of the $V^\ast\to Z\gamma$ ($V=Z$, $\gamma$) process, which allows us to define a new type of left-right asymmetry. In Sec. \ref{numana}, we obtain the bounds on FCNCs mediated by the $Z$ boson to assess their contributions to the $h_3^Z$ form factor. Additionally, we numerically examine the behavior of the polarized partial widths and the asymmetry. Finally, we present the conclusions and prospects in Sec. \ref{final}.

\section{Trilinear neutral gauge boson couplings and FCNC couplings of the top quark}\label{teo}
In this section, we discuss some aspects of the theoretical framework of TNGBCs and the top quark FCNC couplings.

\subsection{$Z\gamma V^\ast$ ($V=Z$, $\gamma$) vertex function}

\begin{figure}[!htb]
\begin{center}
\includegraphics[width=.7\textwidth]{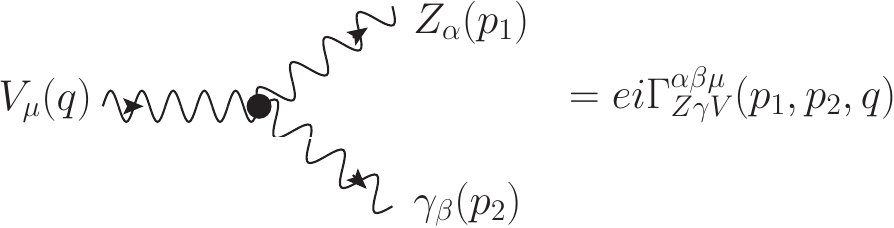}
\caption{Kinematics for the TNGBCs $Z\gamma V^\ast$ ($V=\gamma$, $Z$)\label{Kin}}
\end{center}
\end{figure}

We are focusing on TNGBCs of the form $Z\gamma V^\ast$ ($V=Z$, $\gamma$), which can emerge from dimension-six and dimension-eight operators that are invariant under the $SU(2)_L\times U(1)_Y$ gauge groupe \cite{Gounaris:2000dn,Larios:2000ni}. According to the kinematics shown in Fig. \ref{Kin}, the vertex function can be parametrized as follows 
\begin{align}
\label{Vertex}
  \Gamma_{Z\gamma V^\ast}^{\alpha\beta\mu}\left(p_1,p_2,q\right)=& \frac{i (q^2-m_V^2)}{m_Z^2}\Big[ h_1^V\left(p_2^\mu g^{\alpha\beta}+p_2^\alpha g^{\mu\beta}\right)  + \frac{h_2^V}{m_Z^2}q^\alpha \left( q\cdot p_2 g^{\mu\beta}-p_2^\mu q^\beta\right)\nonumber \\
    &-h_3^V \epsilon^{\mu\alpha\beta\rho} p_{2_\rho}-\frac{h^V_4}{m_Z^2}q^\alpha \epsilon^{\mu\beta\rho\sigma}q_\rho p_{2_\sigma}\Big],
\end{align}
where  $h_3^V$ and $h_4^V$ are $CP$-conserving form factors, whereas $h_1^V$ and $h_2^V$ are $CP$-violating.  In the SM,  only $h_3^V$ ($V=\gamma, Z$) is induced at the one-loop level via fermion exchange, with the dominant contribution being of the order of $10^{-4}$ \cite{Gounaris:2000tb}. As far as new physics contributions are concerned, the $CP$ violating form factor $h_1^Z$ can be induced through a charged scalar loop  \cite{Moyotl:2015bia}, while in Ref. \cite{Hernandez-Juarez:2021mhi} they arise from FCNCs mediated by the $Z$ gauge boson for the $ZZV^\ast$ case.

The ATLAS collaboration obtained the current limits on the $CP$-conserving anomalous couplings of the $Z\gamma V^\ast$ vertex by analyzing the rate and kinematic properties of $Z\gamma$ production at $\sqrt{s}=$ 13 TeV,  with a confidence level of 95\%  \cite{ATLAS:2018nci}:

\begin{align}
    &-3.7 \times10^{-4} < h_3^\gamma < 3.7\times 10^{-4}, \\
    &-4.4 \times10^{-7}< h_4^\gamma <4.3 \times10^{-7},\\
     &-3.2 \times10^{-4}< h_3^Z < 3.3\times10^{-4},\\
    &-4.5\times10^{-7}< h_4^Z <4.4 \times10^{-7}.
\end{align}
This study was conducted only in terms of the $CP$-conserving form factors, whereas similar sensitivity is expected for the $CP$-violating ones \cite{Baur:1992cd}. 

In a particular theory, the $h_i^V$  form factors can be explicitly calculated as functions of the off-shell boson squared four-momentum $q^2$  \cite{Hernandez-Juarez:2020drn,Hernandez-Juarez:2020xon,Hernandez-Juarez:2021xhy}, giving rise to absorptive parts that can be derived either using the Cutkosky rules or through conventional diagrammatic calculations \cite{Cutkosky_1960,Zhou:2004gm}. It turns out that the magnitude of the imaginary parts of the $h_i^V$ form factors can be larger than the real ones \cite{Gounaris:2000tb}. 

\subsection{FCNC couplings of the top quark}

The relevant contributions to the TNGBCs from FCNC are expected to arise from the couplings $Z\overline{t}q$ ($q=c$, $u$) \cite{Hernandez-Juarez:2021mhi}. Within the SM, these couplings are highly suppressed, with the branching ratio $\mathcal{B}\big(t\rightarrow Zc \big)$ being of the order of $10^{-14}$, whereas $\mathcal{B}\big(t\rightarrow Zu \big)$ is approximately three orders of magnitude smaller  \cite{Aguilar-Saavedra:2004mfd}. At the LHC, the ATLAS and CMS collaborations have studied top quark FCNC processes through various decay modes, including $t\rightarrow j\ell^+\ell^-$ \cite{ATLAS:2012hfh}, $t\rightarrow Hq$ \cite{ATLAS:2016qxw,CMS:2017bhz,ATLAS:2017tas,CMS:2021cqc}, $t\rightarrow gq$ \cite{ATLAS:2015iqc,CMS:2016uzc,ATLAS:2021amo}, $t \rightarrow \gamma q$ \cite{CMS:2015kek,ATLAS:2019mke} and $t\rightarrow Zq$ \cite{ATLAS:2018zsq,ATLAS:2019pcn,ATLAS:2023qzr}. Additionally, these processes have been examined at future colliders, such as the FCC-he \cite{Behera:2020ino,Liu:2020bem,Liu:2020kxt}, CEPC \cite{Shi:2019epw} and lepton-hadron colliders \cite{Alici:2019asv,Khatibi:2021phr}. 

The effective Lagrangian that leads to FCNC top quark couplings can be parametrized as follows \cite{Durieux:2014xla}
  \begin{equation}\begin{aligned}
\label{LagZ}
    \mathcal{L}_{tqZ}=&-\frac{e}{2 s_W c_W}\overline{t}\gamma_\mu\left( {g^{tq}_{V}}-\gamma^5{g^{tq}_{A}} \right)q\ Z^\mu, 
      \end{aligned}
      \end{equation}
where $q=c$, $u$ and $g^{tq}_{V}$ and $g^{tq}_{A}$ are the non-diagonal vector and axial couplings, respectively. While these parameters are generally assumed to be real in effective field theories, we will explore a more general scenario in which $g^{tq}_{V}$ and $g^{tq}_{A}$ are complex. Effective extensions of the SM involving complex couplings have previously been proposed to explain $R_K$ and $R_{K^\ast0}$ anomalies \cite{Alda:2018mfy}, as well as in the analysis of $b\rightarrow s\ell\ell$ transitions \cite{Biswas:2020uaq,DiLuzio:2019jyq,Bissmann:2019gfc}.

The ATLAS collaboration reported stringent constraints on the left-handed (LH) and right-handed (RH) $Z\overline{t}q$  couplings \cite{ATLAS:2023qzr}.
From Lagrangian \eqref{LagZ}, the $t\rightarrow Zq$ branching ratio considering the chirality of the quarks can be written as

\begin{equation}
\label{width}
\Gamma_{t\rightarrow Zq}=\frac{N_f e^2 m_t^3}{2(64) \pi  c_W^2  m_Z^2 s_W^2}f^2_{L, R} \left(1-\frac{m_Z^2}{m_t^2}\right){}^2
   \left(1+2 \frac{m_Z^2}{m_t^2}\right),
    \end{equation} 
with $f_{L, R}$ the LH and RH couplings in terms of $g_{\text{V}}^{tq} $ and $g_{\text{A}}^{tq} $ given as follows
\begin{equation}
\label{chiralcou}
f^2_{L,R}={ |g_{\text{A}}^{tq} }|^2\pm 2 |g_{\text{A}}^{tq}|| g_{\text{V}}^{tq}|\cos{\theta}+|{g_{\text{V}}^{tq} }|^2,
    \end{equation} 
where $\theta$ holds for the sum of the $g_{\text{V}}^{tq}$ and $g_{\text{A}}^{tq}$ complex phases, and we have neglected the masses of the light quarks.


\section{FCNC contribution to TNGBCs and the $V^\ast\to Z\gamma$ ($V=Z$, $\gamma$) process}\label{analyticSec}

In this section, we compute the FCNC contributions from the Lagrangian \eqref{LagZ} to the $Z\gamma V^\ast$ vertex ($V=Z$, $\gamma$). Furthermore, we analyze the unpolarized and polarized amplitudes of the $V^\ast\to Z\gamma$ ($V=Z$, $\gamma$) process.

\subsection{FCNC contribution to $h_3^Z$}

\begin{figure}[H]
\begin{center}
\includegraphics[width=.5\textwidth]{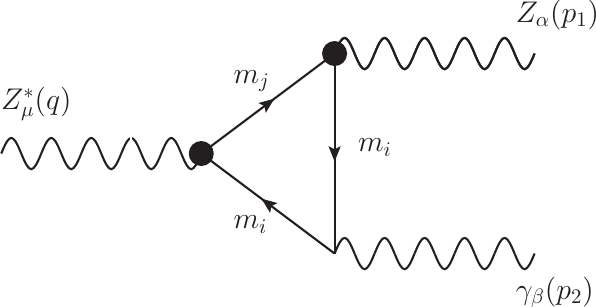}
\caption{Generic Feynman diagram for the contribution of FCNC interaction (denoted by the large dots) to the  $Z\gamma Z^\ast$ vertex at the one-loop level. \label{FeynDiag}}
\end{center}
\end{figure}

In Fig. \ref{FeynDiag}, we show the generic one-loop contribution to the $Z\gamma Z^\ast$ vertex resulting from FCNC couplings. By considering the corresponding permutations, we solved the loop integrals using the Passarino-Veltman reduction scheme with the help of FeynCalc \cite{Shtabovenko:2020gxv}. Furthermore, we adopted the Breitenlohner-Maison-t'Hooft-Veltman (BMHV) scheme for the treatment of the $\gamma_5$ matrix in $d$ dimensions. It is noteworthy that the BMHV scheme can occasionally lead to non-conservation of currents in the final results \cite{BARROSO1991123,Jegerlehner:2001tc}. However, we have verified that currents are conserved in our calculation without necessitating any finite counterterms. For comparative analysis, we also performed an alternative calculation with Package-X \cite{Patel:2015tea}. It is important to note that the $Z\gamma\gamma^\ast$ case cannot be generated through FCNC couplings, as we consider the photon couples to a conserved current.

 In contrast to the $ZZV^\ast$ case, we find that the $Z\gamma V^\ast$ vertex does not give rise to $CP$ violation through complex FCNC couplings. However,  it is possible to obtain a new contribution to the $CP$-conserving form factor $h_3^Z$, which is expressed as follows:
\begin{align}
\label{h3}
h_3^Z=&\sum_i\sum_{i\neq j}\frac{N_f \mathcal{Q}_i e^2 m_Z^2 {\rm Re}\big({g^{ij\ast}_{A}}g^{ij}_{V}\big)}{4 \pi 
   ^2  s_W^2c_W^2\left(m_Z^2-Q^2\right)^3}\Bigg\{2 m_Z^2
   Q^2
  \Big[ \text{B}_0\left(m_Z^2,m_i^2,m_j^2\right)-
   \text{B}_0\left(Q^2,m_i^2,m_j^2\right)\Big]\nonumber\\
   &+\left(m_Z
   ^4-Q^4\right) \Big[ m_i^2
   \text{C}_0\left(m_Z^2,0,Q^2
   ,m_j^2,m_i^2,m_i^2\right)+m_j^2
   \text{C}_0\left(m_Z^2,0,Q^2
   ,m_i^2,m_j^2,m_j^2\right)+1\Big]\Bigg\},
\end{align}
where $N_f$, $m_i$, and $\mathcal{Q}_i$ are the color number, mass, and electric charge of the fermion $m_i$, respectively. The notation $\sqrt{q^2}\equiv Q$ has also been introduced. It is important to note that $h_3^Z$ is free of ultraviolet divergences and is consistent with the result for real FCNC couplings reported in Ref. \refcite{Gounaris:2000tb}. 

In the high-energy limit $Q^2\gg m_i^2$, $m_j^2$, $m_Z^2$, Eq. \eqref{h3} reduces to 
\begin{equation}
\label{h3l}
h_3^Z\approx \sum_i\sum_{i\neq j}\frac{N_f \mathcal{Q}_i e^2 m_Z^2 {\rm Re}\big({g^{ij\ast}_{A}}g^{ij}_{V}\big)}{4 \pi 
   ^2  s_W^2c_W^2 Q^2},
\end{equation}
this result agrees with the flavor-conserving case ($m_i=m_j$), although it is twice the expression reported in \cite{Gounaris:2000tb}. This discrepancy arises because we are considering additional Feynman diagrams resulting from the fermion exchange in the loop. We observe that the anomalous coupling $h^Z_3$ approaches 0 for high values of $Q$.

\subsection{Unpolarized  $V^*\to Z\gamma$ ($V=Z,\gamma$) decay widths}\label{secUn}

We calculate the unpolarized partial widths for the process $V^\ast\to Z\gamma$, which will be essential for cross-checking our polarized results. The amplitudes for an off-shell $Z$ or photon are computed separately. We do not consider the interference terms between these amplitudes, as their treatment differs significantly in the collider process $f\overline{f}\to V^\ast \to Z\gamma$, where the off-shell boson acts as a propagator.   

Using the vertex function in Eq. \eqref{Vertex}  and considering complex form factors $h_i^V={\rm Re}\big[h_i^V\big]+ i\ {\rm Im}\big[h_i^V\big]$ ($i=1,\ldots,4$),
we obtain the following amplitude for the production of a $Z\gamma$ pair from an off-shell $V$ gauge boson 
\begin{align}
\label{amplitude}
  \mathcal{M}(V^*\to Z\gamma)  =& -e\frac{(q^2-m_V^2)}{m_Z^2}\Bigg[ \big({\rm Re}\big[h_1^V\big]+ i\ {\rm Im}\big[h_1^V\big]\big)\left(p_2^\mu g^{\alpha\beta}+p_2^\alpha g^{\mu\beta}\right) \nonumber\\
  & + \frac{{\rm Re}\big[h_2^V\big]+ i\ {\rm Im}\big[h_2^V\big]}{m_Z^2}q^\alpha \left( q\cdot p_2 g^{\mu\beta}-p_2^\mu q^\beta\right)\nonumber \\
    &-\big({\rm Re}\big[h_3^V\big]+ i\ {\rm Im}\big[h_3^V\big]\big)^V \epsilon^{\mu\alpha\beta\rho} p_{2_\rho}-\frac{{\rm Re}\big[h_4^V\big]+ i\ {\rm Im}\big[h_4^V\big]}{m_Z^2}q^\alpha \epsilon^{\mu\beta\rho\sigma}q_\rho p_{2_\sigma}\Bigg]\nonumber\\
    &\times\epsilon^\ast(p_1,\lambda_1)\epsilon^\ast(p_2,\lambda_2)\epsilon(q,\lambda_3),
\end{align}
where $\lambda$ and $\epsilon(r,\lambda_i)$ (i=1,2,3) correspond to the polarizations and polarization vector of the neutral gauge bosons, respectively. 

By using standard techniques to calculate Feynman diagrams, we can obtain from Eq. \eqref{amplitude} the unpolarized partial decay width $\Gamma(V^\ast\rightarrow Z \gamma)$ ($V=Z$, $\gamma$). For the scenario $V^\ast=Z$, it can be expressed as follows
\begin{align}\label{ZunpolG}
\Gamma_{Z^\ast\rightarrow Z\gamma}=&\frac{e^2
   \left(m_Z-Q\right){}^4
   \left(m_Z+Q\right){}^4
   \left(Q^2-m_Z^2\right)}{128 \pi 
   Q^3 m_Z^{10}}
   \Bigg\{-2 Q^2 m_Z^2
   \Big({\rm Re}\big[h_2^Z\big]{}^2+{\rm Re}\big[h_4^Z\big]{}^2+2\Big\{
  {\rm Re}\big[h_1^Z\big]
   {\rm Re}\big[h_2^Z\big]\nonumber\\
   & +
   {\rm Re}\big[h_3^Z\big]
   {\rm Re}\big[h_4^Z\big]+
   {\rm Im}\big[h_1^Z\big]
   {\rm Im}\big[h_2^Z\big]+ {\rm Im}\big[h_3^Z\big]
   {\rm Im}\big[h_4^Z\big]
  \Big\}+{\rm Im}\big[h_2^Z\big]
   {}^2+{\rm Im}\big[h_4^Z\big]{}^2\Big)\nonumber\\
   &+m_Z^4 \Big({\rm Re}\big[h_2^Z\big]{}^2+{\rm Re}\big[h_4^Z\big]{}^2+4\Big\{
   2
   {\rm Re}\big[h_1^Z\big]{}^2+2
   {\rm Re}\big[h_3^Z\big]{}^2+2{\rm Im}\big[h_1^Z\big]{}^2+2
   {\rm Im}\big[h_3^Z\big]{}^2+
   {\rm Re}\big[h_1^Z\big]
   {\rm Re}\big[h_2^Z\big]\nonumber\\ &
   +
   {\rm Re}\big[h_3^Z\big]
   {\rm Re}\big[h_4^Z\big]+
   {\rm Im}\big[h_2^Z\big]
   {\rm Im}\big[h_1^Z\big]+
   {\rm Im}\big[h_3^Z\big]
   {\rm Im}\big[h_4^Z\big]\Big\}+{\rm Im}\big[h_2^Z\big]{}^2+{\rm Im}\big[h_4^Z\big]{}^2\Big)\nonumber\\&
   +Q^4
   \Big({\rm Re}\big[h_2^Z\big]{}
   ^2+{\rm Re}\big[h_4^Z\big]{}^
   2+{\rm Im}\big[h_2^Z\big]
   {}^2+{\rm Im}\big[h_4^Z\big]{
   }^2\Big)\Bigg\},
\end{align}
whereas for an off-shell photon, we obtain
\begin{align}\label{PhotonunpolG}
\Gamma_{\gamma^\ast\rightarrow Z\gamma}=&\frac{e^2 Q
   \left(Q^2-m_Z^2\right){}^3}
   {128 \pi  m_Z^{10}}  
   \Bigg\{-2 Q^2 m_Z^2
   \Big({\rm Re}\big[h_2^{\gamma }\big]{}^2+{\rm Re}\big[h_4^{\gamma }\big]{}^2+
   2\Big\{
   {\rm Re}\big[h_1^{\gamma }\big]
   {\rm Re}\big[h_2^{\gamma }\big]+
   {\rm Re}\big[h_3^{\gamma }\big]
   {\rm Re}\big[h_4^{\gamma }\big]\nonumber\\
   &
   +
   {\rm Im}\big[h_1^{\gamma }\big]
   {\rm Im}\big[h_2^{\gamma }\big]+
   {\rm Im}\big[h_3^{\gamma }\big]
   {\rm Im}\big[h_4^{\gamma }\big]\Big\}+{\rm Im}\big[h_2^{\gamma }\big]{}^2+{\rm Im}\big[h_4^{\gamma }\big]{}^2\Big)+m_Z^4
   \Big(\Big\{2
   {\rm Im}\big[h_1^{\gamma }\big]+{\rm Im}\big[h_2^{\gamma }\big]\Big\}{}^2\nonumber\\
   &+\Big\{2
   {\rm Im}\big[h_3^{\gamma }\big]+{\rm Im}\big[h_4^{\gamma }\big]\Big\}{}^2+\Big\{2
   {\rm Re}\big[h_1^{\gamma }\big]+{\rm Re}\big[h_2^{\gamma }\big]\Big\}{}^2+\Big\{2
   {\rm Re}\big[h_3^{\gamma }\big]+{\rm Re}\big[h_4^{\gamma }\big]\Big\}{}^2\Big)\nonumber\\
   &+
   Q^4
   \Big({\rm Re}\big[h_2^{\gamma }\big]{}^2+{\rm Re}\big[h_4^{\gamma }\big]{}^2+{\rm Im}\big[h_2^{\gamma }\big]{}^2+{\rm Im}\big[h_4^{\gamma }\big]{}^2\Big)\Bigg\}
    \end{align}
We have used the notation $Q=\sqrt{q^2}$  to denote the norm of the four-momentum of the off-shell gauge boson $V$. As expected, there are no interference terms between the $CP$-violating and $CP$-conserving form factors \cite{Baur:1992cd}. 

In the calculation of the squared amplitudes for the processes $V^\ast\to Z\gamma$ ($V=Z$, $\gamma$), it is crucial to average over the polarizations of the decaying particle. However, in the context of a collider process, the off-shell $V^\ast$ boson acts as a propagator within the amplitude. For instance,  consider the process $e^+e^-\to V^\ast\to Z\gamma$ in an electron-positron collider. In this scenario, the $V^\ast$ ($V=Z$, $\gamma$) boson is not part of the initial state, so it is unnecessary to average over its polarizations. Consequently, in our results presented in Eqs. \eqref{ZunpolG} and \eqref{PhotonunpolG}, we have omitted a factor of 1/3 and 1/2, respectively.

\subsection{Polarized partial decay widths}\label{secPol}

 We will present analytical expressions for the polarized partial decay widths of the process $V^\ast\rightarrow Z\gamma$. The polarization of $Z$ gauge bosons has been studied at the LHC through leptonic decays \cite{ATLAS:2023zrv,ATLAS:2016rnf,ATLAS:2023lsr,CMS:2015cyj,LHCb:2022tbc}. Similarly, the polarizations of photons have been investigated at the LHC, Belle, and BaBar \cite{LHCb:2019vks,LHCb:2021byf,Belle:2006pxp,BaBar:2008okc}. 
While, the polarizations of  TNGBCs have been studied in the context of polarized beams \cite{Rizzo:1999xj,Atag:2003wm,Ananthanarayan:2004eb,Ananthanarayan:2011fr,Ananthanarayan:2014sea,Rahaman:2017qql,Subba:2023jia} and also considering the polarizations of the final state gauge bosons \cite{Atag:2004cn,Rahaman:2016pqj,Rahaman:2018ujg}.

From Eq. \eqref{amplitude} and following the approach in Ref. \cite{Hernandez-Juarez:2023dor}, we can also calculate the polarized partial decay widths $ \Gamma^{\lambda}(V^\ast\rightarrow Z\gamma)$, where $\lambda$ is the polarization of the $Z$ gauge boson.
 In a frame where the off-shell gauge boson $V^\ast$ is at rest and the $Z$ boson moves along the positive $x$ axis, the polarization vectors are
\begin{equation}
\epsilon(p_1,0)=\frac{1}{2 m_Z Q}(m_Z^2-Q^2,m_Z^2+Q^2,0,0),
\end{equation}
\begin{equation}
\epsilon(p_1,L/R)=\frac{1}{\sqrt{2}}(0,0,-i,\mp1),
\end{equation}
where $L$ and $R$ denote the left and right polarizations, whereas $0$ corresponds to the longitudinal polarization. In our calculation, we restrict our attention to the polarizations of the $Z$ gauge boson. Moreover, analogous to the treatment in the unpolarized case, we do not average over initial polarizations.  

The polarized partial decay widths involving a transversely polarized $Z$ boson is only found for the case of an off-shell $Z$ boson and are given as 
\begin{align}\label{polZLL}
\Gamma^{L,R}(Z^\ast\rightarrow Z\gamma)=&\frac{e^2
   \left(Q^2-m_Z^2\right)^5}{6
   4 \pi  Q^3 m_Z^6}
   \Big\{{\rm Re}\big[h_1^Z\big]{}^2+{\rm Re}\big[h_3^Z\big]{}^2\pm 2\Big(
   {\rm Im}\big[h_1^Z\big]
   {\rm Re}\big[h_3^Z\big]-
   {\rm Im}\big[h_3^Z\big]
   {\rm Re}\big[h_1^Z\big]\Big)\nonumber\\&
   +{\rm Im}\big[h_1^Z\big]{}^2+{\rm Im}\big[h_3^Z\big]{}^2\Big\},
   \end{align}
where the $+(-)$ sign is for $L$ $(R)$. For a longitudinally polarized $Z$ boson and an off-shell $Z$ boson, the partial widths are expressed as follows
\begin{align}\label{polZ0L}
\Gamma^{0}(Z^\ast\rightarrow Z\gamma)=&\frac{e^2\left(Q^2-m_Z^2\right)^5}{128 \pi  Q^3 m_Z^{10}}
   \Bigg\{-2 Q^2 m_Z^2 \Big({\rm Re}\big[h_2^Z\big]^2+{\rm Im}\big[h_2^Z\big]^2+2\Big\{ {\rm Re}\big[h_1^Z\big]{\rm Re}\big[h_2^Z\big]+{\rm Im}\big[h_1^Z\big]{\rm Im}\big[h_2^Z\big]   \nonumber\\
   &+{\rm Re}\big[h_3^Z\big]{\rm Re}\big[h_4^Z\big]+{\rm Im}\big[h_3^Z\big]{\rm Im}\big[h_4^Z\big]\Big\}+{\rm Re}\big[h_4^Z\big]^2+{\rm Im}\big[h_4^Z\big]^2\Big)\nonumber\\
   &+m_Z^4 \Big(\Big\{2
   {\rm Re}\big[h_1^Z\big]+{\rm Re}\big[h_2^Z\big]\Big\}^2
   +\Big\{2
   {\rm Im}\big[h_1^Z\big]+{\rm Im}\big[h_2^Z\big]\Big\} {}^2
   +\Big\{2
   {\rm Re}\big[h_3^Z\big]+{\rm Re}\big[h_4^Z\big]\Big\}^2\nonumber\\
   &
   +\Big\{2
   {\rm Im}\big[h_3^Z\big]+{\rm Im}\big[h_4^Z\big]\Big\}{}^2\Big)
    +Q^4
   \Big({\rm Re}\big[h_2^Z\big]^2+{\rm Im}\left[h
   _2^Z\right]^2+ {\rm Re}\big[h_4^Z\big]^2+{\rm Im}\big[h_4^Z\big]^2\Big)\Bigg\},
\end{align}
whereas $\Gamma^{0R}(Z^\ast\rightarrow Z\gamma)$ is obtained from \eqref{polZ0L} after the replacement $h_{3,4}^Z\to -h_{3,4}^Z$.

In the SM, where only  $h_3^Z$ is generated at the one-loop level, the above expressions reduce to
 \begin{equation}
\label{polSM}
\Gamma^{\lambda}_{SM}(Z^\ast\rightarrow Z\gamma)=\frac{e^2 (Q^2-m_Z^2)^5 \kappa_\lambda}{64\pi m_Z^6 Q^3}\Big\{ {\rm Re}\big[h_3^Z\big]^2+{\rm Im}\big[h_3^Z\big] ^2\Big\}\text{,}\quad \lambda=0\text{, } L\text{, }R,
\end{equation}
with $\kappa_\lambda=1$ for the $L$ and $R$ polarizations and $\kappa=2$ for the 0 polarization. To cross-check our results, we verified that the expression in Eq. \eqref{ZunpolG} is obtained by summing over all the polarized partial widths. 

For the $\gamma^\ast\rightarrow Z\gamma$ process, only a longitudinally polarized $Z$ boson can be produced in the final state. The polarized partial decay width can be written in terms of  $\Gamma^{0}(Z^\ast\rightarrow Z\gamma)$ as follows 
   \begin{align}\label{polGamma}
\Gamma^{0}(\gamma^\ast\rightarrow Z\gamma)=\frac{Q^4}{(Q^2-m_Z^2)^2}\Gamma^{0}(Z^\ast\rightarrow Z\gamma).
   \end{align}
This equation is the same as that obtained in Eq. \ref{PhotonunpolG}. Thus, we do not anticipate relevant results for polarized processes when examining an off-shell photon in the $V^\ast Z\gamma$ vertex. Measuring the photon polarization is not feasible at colliders and has not been addressed in this discussion. However, some interesting results can emerge when considering polarized photons in combination with a longitudinally polarized $Z$ boson. These results can be found in  \ref{AppG}.

It is important to emphasize several notable aspects of the results presented above.
The transversely polarized partial widths exhibit different sensitivity to $CP$-conserving and $CP$-violating form factors compared to the unpolarized case \cite{Baur:1992cd}. Therefore, these observables have the potential to distinguish new physics contributions. Furthermore, the $\Gamma^{L, R}$ partial widths arise specifically from an off-shell $Z$ boson. This result indicates we can identify the contributions from the $Z\gamma Z^\ast$ vertex in $V^\ast\to Z\gamma$ production by analyzing transversely polarized $Z$ bosons. In contrast, such an analysis would be significantly more intricate for unpolarized gauge bosons in the final state,  requiring certain assumptions regarding the values of the remaining $h_i^V$ ($V=Z$, $\gamma$) form factors  \cite{ATLAS:2018nci}. Finally, unlike the unpolarized case, the polarized decay widths include contributions from interference terms between the $CP$-conserving and $CP$-violating form factors. These terms change only by a sign for left and right polarizations, allowing for the definition of asymmetries that can help to disentangle the effects of the $CP$-violating form factors associated with TNGBCs.

\subsection{The left-right asymmetry}\label{Assubsec}

Interesting observables, such as left-right asymmetries, can be defined as a result of the disparity in sign between the interference terms in Eq. \eqref{polZLL}. For the case of the $Z^\ast\rightarrow Z\gamma$ process, we can define the $\mathcal{A}_{LR}$ asymmetry as follows 
 \begin{equation}
\label{ALR1}
\mathcal{A}_{LR}=\frac{\Gamma^{L}(Z^\ast\rightarrow Z\gamma)-\Gamma^{R}(Z^\ast\rightarrow Z\gamma)}{\Gamma^{L}(Z^\ast\rightarrow Z\gamma)+\Gamma^{R}(Z^\ast\rightarrow Z\gamma)},
\end{equation}   
which, after the replacement of the polarized decay widths, reads
   \begin{align}\label{ALR11}
\mathcal{A}_{LR}=\frac{2
   \left({\rm Im}\big[h_1^Z\big]
   {\rm Re}\big[h_3^Z\big]-{\rm Im}\big[h_3^Z\big]
   {\rm Re}\big[h_1^Z\big]\right)}{{\rm Im}\big[h_1^Z\big]{}^
   2+{\rm Im}\big[h_3^Z\big]{}^2
   +{\rm Re}\big[h_1^Z\big]{}^2+
   {\rm Re}\big[h_3^Z\big]{}^2}.
\end{align}
We observe in Eq. \eqref{ALR11} that a non-vanishing $\mathcal{A}_{LR}$  requires the $CP$-violating form factor $h_1^Z$, as well as the imaginary parts of the $h_1^Z$ and $h_3^Z$ form factors. It is well established in the SM that the $h_3^Z$ form factor is complex, while  $h_1^Z$ is not induced \cite{Gounaris:2000tb}. As a result, the $\mathcal{A}_{LR}$ asymmetry vanishes. On the other hand, a non-zero $\mathcal{A}_{LR}$  would suggest the existence of $CP$ violation and could provide valuable constraints on $h_1^Z$. 

 In the context of TNGBCs, angular asymmetries have been discussed in Refs. \cite{Rizzo:1999xj,Ananthanarayan:2004eb,Ananthanarayan:2014sea,Rahaman:2016pqj,Rahaman:2017qql,Rahaman:2018ujg}. Furthermore, it has been pointed out that these observables could provide limits on TNGBCs that may be comparable to the current limits established by the LHC \cite{Rahaman:2018ujg}.

\section{Numerical results}\label{numana}

In this section, we obtain numerical values for the contribution from FCNC couplings to $h_3^Z$ in Eq. \eqref{h3}, as well as the polarized partial widths. We expect that the dominant contributions to $h_3^Z$ will arise from the FCNC couplings of the top quark \cite{Hernandez-Juarez:2021mhi}. Therefore, we begin by establishing constraints on these couplings.

\subsection{Constraints FCNC couplings of the top quark}

\begin{table}[!tbh]
\caption{Current upper bounds with 95\% CL on LH and RH couplings of the $t\to qZ$ decays in units of $10^{-4}$. The limits at the first two rows apply for both LH and RH couplings. \label{FCNCbounds}}
\begin{center}
\begin{tabular}{ccc}
\hline
\hline
$\mathcal{B}(t\rightarrow qZ)$&Coupling&Observed at 95\% CL\\\hline\hline
$q=u$&LH/RH&$1.6$\\
$q=c$&LH/RH&$2.4$\\
$q=u$&LH&$0.62$\\
$q=u$&RH&$0.66$\\
$q=c$&LH&$1.3$\\
$q=c$&RH&$1.2$\\
\hline
\hline
\end{tabular}
\end{center}
\end{table}

\begin{figure}[t]
\centerline{\includegraphics[width=.5\textwidth]{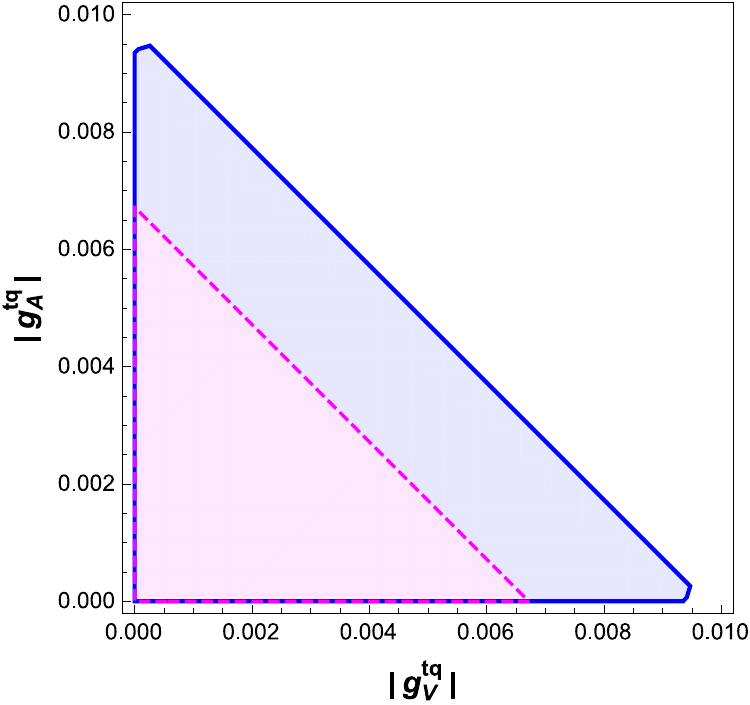}}
\caption{Allowed areas with 95\% confidence-level in the $|g_{\text{V}}^{tq}|$  vs $|g_{\text{A}}^{tq}|$ plane from the current bounds on the branching ratio $t\rightarrow Zq$  for left- and right-handed quarks. The solid-line (dashed-line) boundaries correspond to the $Z\overline{t}c$ ($Z\overline{t}u$) couplings. We have considered $\theta=0$.}\label{Plot1}
\end{figure}

We now turn our attention to constraining the  ${g^{ij}_{A}}$ and $g^{ij}_{V}$ couplings of the top quark to estimate the FCNC contributions to $h_3^Z$. The stringent limits on the branching ratios for the $  t\rightarrow qZ$ decay, with a confidence level of 95\%, have been reported by the ATLAS collaboration for LH and RH quarks \cite{ATLAS:2023qzr}. These limits are presented in Table \ref{FCNCbounds} for convenience.
 By using numerical values for the masses and constants in Eq. \eqref{width}, the $t\to Zq$ branching ratio considering the chirality of the quarks can be expressed as follows
\begin{equation}
\label{branch2}
\mathcal{B}(t\rightarrow qZ)=1.37355 \Big( |{g_{\text{A}}^{tq} }|^2\pm 2 |g_{\text{A}}^{tq}|| g_{\text{V}}^{tq}|\cos{\theta}+|{g_{\text{V}}^{tq} }|^2\Big).
\end{equation}
 From Eq. \eqref{branch2}, it is possible to delineate the permissible region in the $|g_{\text{V}}^{tq}|$ vs $|g_{\text{A}}^{tq}|$ plane.  In  Fig. \ref{Plot1}, we show the allowed values for the vector and axial couplings, where the most stringent constraints arising from the limits on LH and RH decays are used \cite{ATLAS:2023qzr}. Our bounds can be summarized as follows
\begin{align}
 |g_{\text{V}}^{tu}|\text{,}|g_{\text{A}}^{tu}|\leqslant 0.007,\quad  |g_{\text{V}}^{tc}|\text{,}|g_{\text{A}}^{tc}|\leqslant 0.0095.   \label{cotas2}
 \end{align}
Additional bounds can be derived from $B$ and $K$ meson decays (for instance: $B\to K^\ast \ell^+\ell^-$ and $K\to\pi\nu\overline{\nu}$) \cite{Li:2011af,Gong:2013sh,Chen:2018lze,Kumbhakar:2019njm}. Nevertheless, to our knowledge, these constraints are of a similar order of magnitude to those presented in this study. For this reason, we will utilize the limits outlined in Eq. \eqref{cotas2}. Bounds on FCNC of the $Z$ gauge boson to down-type quarks have been established from the decays of $K$ and $B$ mesons \cite{Buchalla:2000sk,Mohanta:2005gm,Silverman:1991fi,Buras:1998ed,Giri:2003jj}, which are of the order $10^{-3}$ for both vector and axial couplings. Moreover, for lepton flavor-violating couplings mediated by the $Z$ gauge boson, constraints can be derived from the decays $\mu\rightarrow eee$,  $\tau^-\rightarrow e^-\mu^+\mu^-$ and $\tau^-\rightarrow \mu^-\mu^+\mu^-$, which lead to limits in the range $10^{-3}-10^{-6}$ \cite{Mohanta:2010yj}. 

\subsection{ FCNC contributions to $h_3^Z$}

We will now analyze the behavior of the FCNC contributions to the $CP$-conserving form factor $h_3^Z$. For the numerical evaluation of the Passarino-Veltman scalar functions, we utilized LoopTools \cite{Hahn:1998yk}.  In Table \ref{TableConstrains}, we present the values of the FCNC couplings used in our numerical analysis. We selected values for the vector and axial couplings of the $Z\overline{t}c$ and $Z\overline{t}u$ interactions that are consistent with the bounds in Fig. \ref{Plot1}. For the remaining FCNC couplings involving up-type quarks, we adopt values of a comparable order of magnitude, while ensuring that the condition $|g_V^{cu}|^2+|g_A^{cu}|^2<|g_V^{tq}|^2+|g_A^{tq}|^2$ ($q=u$, $c$) is satisfied, maintaining the hierarchy of the couplings observed in Fig. \ref{Plot1}. Additionally, we include the FCNC of the $Z$ gauge boson with down-type quarks for completeness. Given that the bounds on $Zd_i \overline{d}_j$ couplings are of order $10^{-3}$, we choose values around this magnitude that are anticipated to contribute significantly to $h_3^Z$. Contributions from the leptons are negligible and thus are not taken into account in our analysis.

\begin{table}[!htb]
\caption{Values for the FCNC couplings of the $Z$ gauge boson to quark pairs considered in our analysis of the FCNC contributions to the $CP$-conserving form factor $h_3^Z$. 
}
\label{TableConstrains}
\begin{center}\setlength{\tabcolsep}{12pt} 
\renewcommand{\arraystretch}{1.5}
\begin{tabular}{ccccc}
\hline
\hline
  &$\overline{t}c$&$\overline{t}u$&$\overline{c}u$&$\overline{d}_id_j$\\
\hline
\hline
 $g^{f_1 f_2}_\text{V}$ &0.009&0.002&$0.001$&$0.001$ \\
 $g^{f_1 f_2}_\text{A}$&0.001&0.005&0.002&0.006\\
\hline
\hline
\end{tabular}
\end{center}
\end{table}

\begin{figure}[!htb] 
\begin{center}
\includegraphics[width=.7\textwidth]{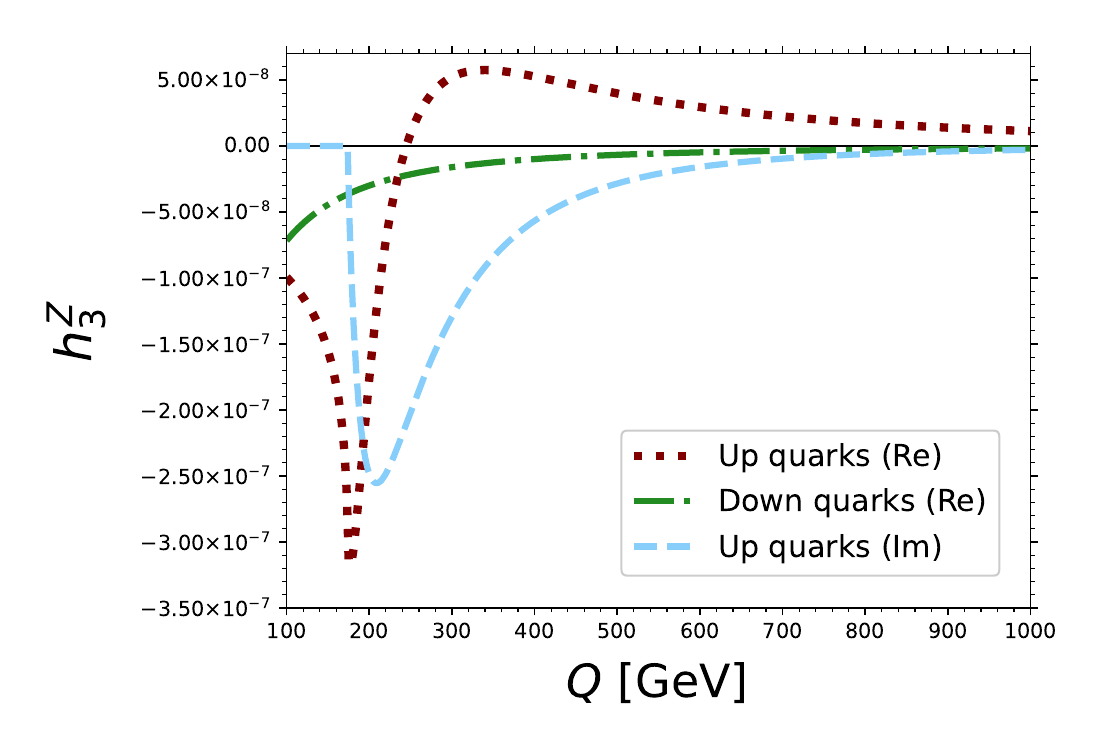}
\caption{Behavior of the FCNC contribution of up- and down-type quarks to the $CP$-conserving form factor $h^Z_3$ as a function of $Q$. We use the values shown in Table \ref{TableConstrains} for the coupling constants. The absorptive part of the down-type quark is negligible and not shown in the plot.\label{h3q}}
\end{center}
\end{figure}

In Fig. \ref{h3q}, we present the total contributions from up-type and down-type quark to $h_3^Z$ as a function of $Q$. As expected, the dominant contributions to the real and imaginary parts of $h_3^Z$ arise from up-type quarks. At low energies, these contributions are of the order of $10^{-7}$, but they diminish rapidly as $Q$ increases. In contrast, the contribution from the down-type quarks to the real part is approximately one order of magnitude smaller than that from the up-type quarks, while the absorptive part is considerably smaller, reaching values around $10^{-11}$, and is not displayed in the figure. We note that the absorptive part of $h_3^Z$ becomes relevant for $Q$ values just above the top quark mass, where the $\overline{t}q$ quark pair in the loop of Fig \ref{FeynDiag} interacting with the off-shell $Z$ gauge boson can be on-shell. Furthermore, in the 200 GeV $\leqslant Q \leqslant$  400 GeV range, we find that the imaginary part of $h_3^Z$ is larger than the real one. Our results are three orders of magnitude smaller than the predictions in the SM and  Minimal Supersymmetric SM  (MSSM), estimated to be around $10^{-4}$ \cite{Gounaris:2000tb}, and remain insufficient to be observed at the LHC. However, these effects could be detected at future colliders, such as the 100 TeV $pp$ collider, which is sensitive to TNGBCs of order $10^{-7}$ \cite{Ellis:2023ucy}. Furthermore, the FCC-hh is anticipated to offer enhanced sensitivity to TNGBCs compared to the LHC \cite{Yilmaz:2021qnv}. On the other hand, in light of the absence of clear signs of superpartners at the LHC \cite{Tata:2020afe}, we have adopted a model-independent approach in our study. This framework allows a modern analysis of the new physics contributions to the $CP$-conserving form factor $h_3^Z$.

\subsection{Behavior of the polarized $V^*\to Z\gamma$ partial decay widths}


The corresponding values for the four scenarios in our analysis are shown in Table \ref{ta2}. We consider the $Z\gamma Z^\ast$ and $Z\gamma\gamma^\ast$ cases to be similar. For the $h_4^V$ form factor, we use values consistent with the current experimental bounds of order $10^{-7}$ \cite{ATLAS:2018nci}. While for $h_3^V(Q)$ ($V=Z$, $\gamma$), we use the SM result reported in Ref. \cite{Gounaris:2000tb}, which depend on $Q$ and are in terms of the Passarino-Veltman scalar functions. In scenarios I and II, we use the same values of order $10^{-4}$ for the $CP$-violating form factors but with differing signs to analyze the effects of negative contributions. In scenario III,  we adopt order $10^{-7}$ values for $h_2^V$ to study the case where the contributions from $h_1^V$ and $h_3^V$ are dominant. Finally, it is anticipated that the $CP$-violating form factors arising from new physics models will depend on $Q$ \cite{Hernandez-Juarez:2021mhi}. Therefore, we consider a more realistic behavior for the form factor $h_1^V$ in scenario IV, which is proportional to $h_3^V(Q)$ from the SM. 

\begin{table}[!htb]\caption{Numerical values for the $h_i^V$ ($V=Z$, $\gamma$ and i=1, 2, 4) form factors considered in our numerical analysis. For $h_3^V$ we use the SM model result.}\label{ta2}
\begin{center}\setlength{\tabcolsep}{12pt} 
\renewcommand{\arraystretch}{1.5}
\begin{tabular}{c ccccc c}\hline\hline
Scenario & ${\rm Re}\big[h^V_1\big] $& ${\rm Im}\big[h^V_1\big] $& ${\rm Re}\big[h^V_2\big]$ & ${\rm Im}\big[h^V_2\big]$&${\rm Re}\big[h^V_4\big]$&${\rm Im}\big[h^V_4\big] $\\\hline\hline 
I & 1$\times10^{-4}$ & -3$\times10^{-4}$ & -1$\times10^{-4}$ & 2$\times10^{-4}$ & 4$\times10^{-7}$& 3$\times10^{-7}$ \\
 II & -1$\times10^{-4}$ & 3$\times10^{-4}$ & 1$\times10^{-4}$ & 2$\times10^{-4}$ &-4$\times10^{-7}$& 3$\times10^{-7}$\\ 
 III & 3$\times10^{-4}$ & 3$\times10^{-4}$ & -1$\times10^{-7}$ & 4$\times10^{-7}$&4$\times10^{-7}$&-3$\times10^{-7}$\\
   IV & 0.1 ${\rm Re}\big[h_3^V\big(Q)]$  & -0.4 ${\rm Im}\big[h_3^V(Q)\big]$ & -1$\times10^{-7}$ & 4$\times10^{-7}$&-3$\times10^{-7}$&1$\times10^{-7}$\\\hline\hline
 \end{tabular}\end{center}\end{table}

\begin{figure}[t]
\begin{center}
\subfigure{}\includegraphics[width=6.25cm]{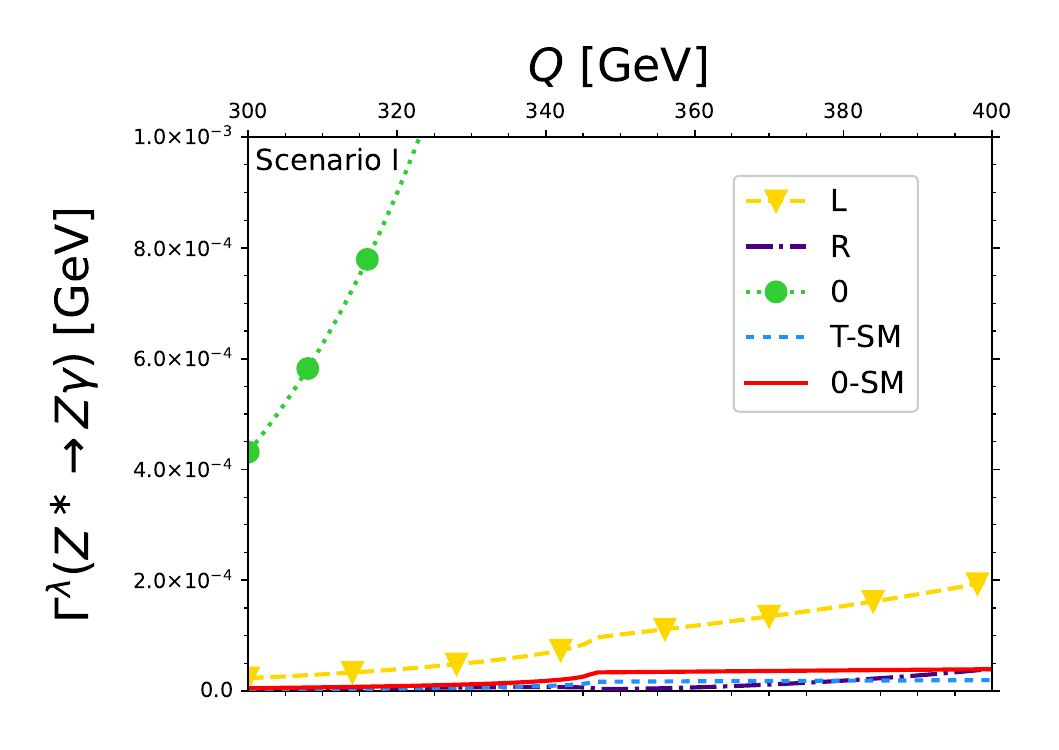}\hspace{-.73cm}
\subfigure{}\includegraphics[width=6.25cm]{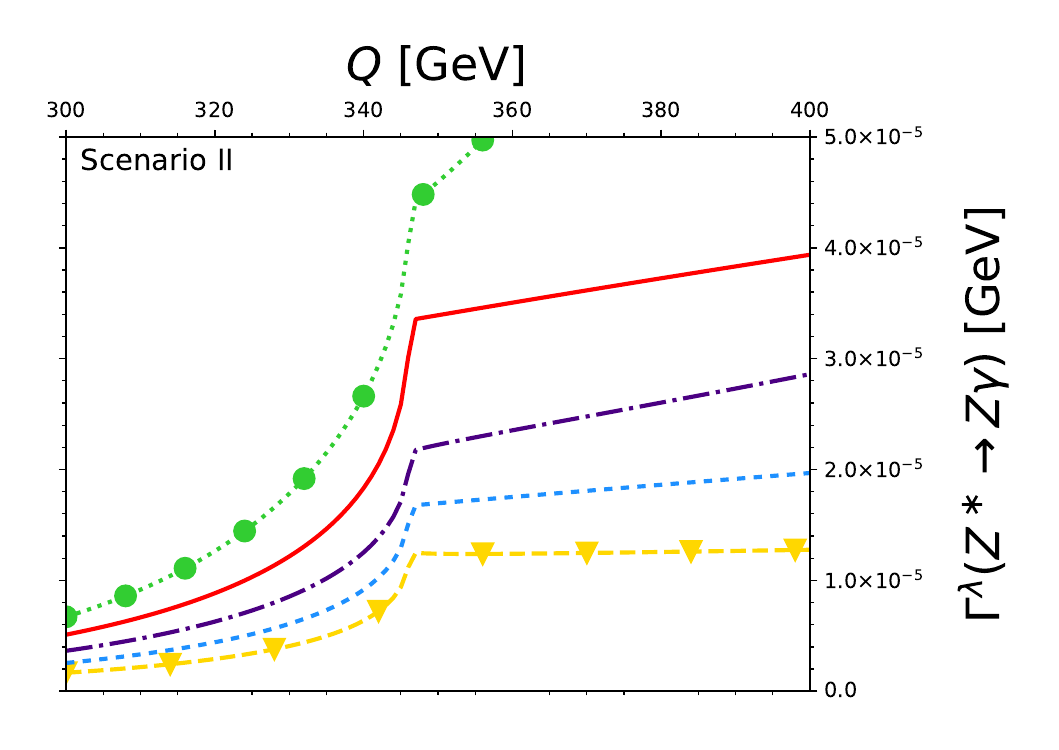}\\\vspace{-.78cm}
\subfigure{}\includegraphics[width=6.25cm]{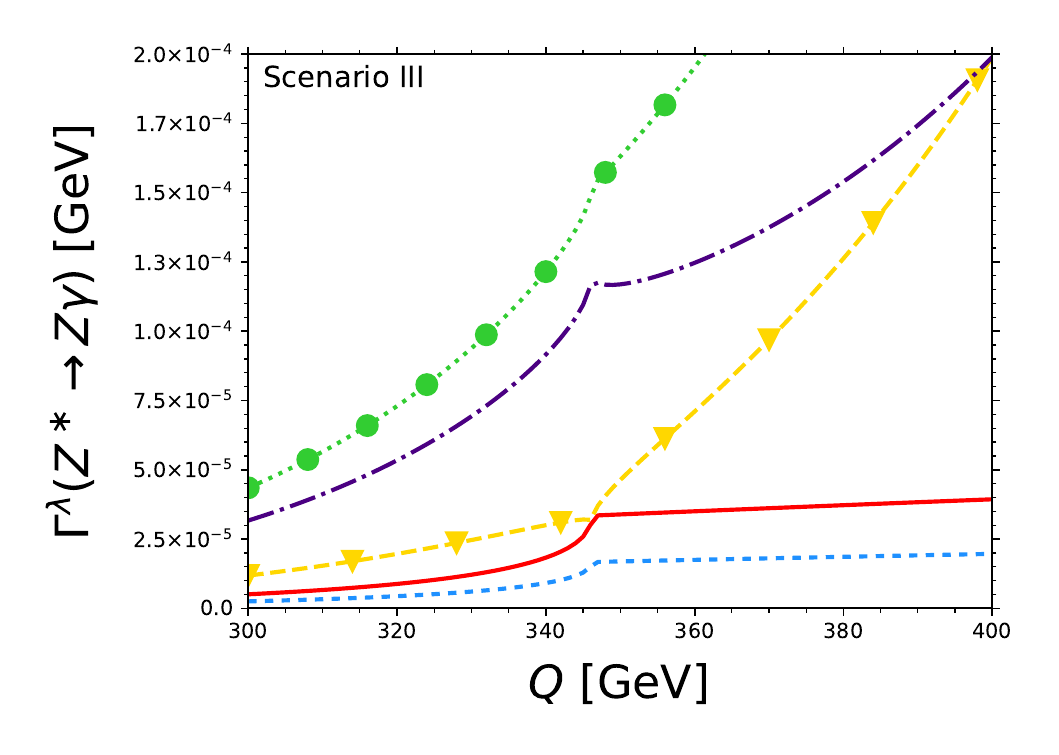}\hspace{-.73cm}
\subfigure{}\includegraphics[width=6.25cm]{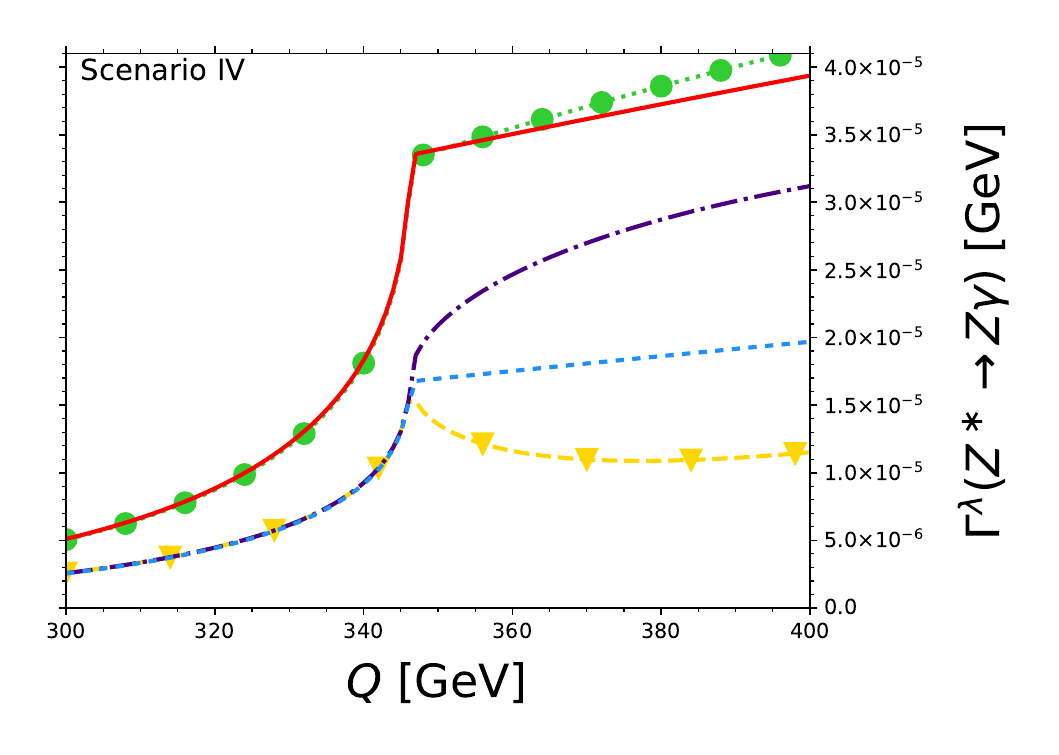}
\caption{Polarized partial  $Z^\ast\rightarrow Z\gamma$ decay width as a function of $Q$ in the four scenarios of Table \ref{ta2} for the values of the form factors. The one-loop level SM contribution is also considered.  \label{PlotsZ1}}
\end{center}
\end{figure}

In Fig. \ref{PlotsZ1}, we show the behavior of the polarized partial decay width for the $Z^\ast\rightarrow Z\gamma$ process as a function of $Q$ across the four scenarios in Table \ref{ta2}. We also include the transversely (T-SM) and longitudinally (0-SM) polarized SM contributions at the one-loop level. In scenario I,  we observe the most significant values, reaching magnitudes of the order of $10^{-3}$ for the $0$ polarization, whereas the remaining fall into the range of $10^{-4}-10^{-5}$, similar to the predictions of the SM. In scenario II, all the polarized partial widths are of order $10^{-5}$, with the longitudinally polarization being the dominant contribution. Scenario III also presents a notable fluctuations for the transversely polarizations from the SM prediction, with all the partial widths reaching values of $10^{-4}$. Significant deviations between the contributions from new physics and the SM are evident in scenarios I-III across the energy range 300 GeV $<Q<$ 400 GeV. In scenario IV, these fluctuations become noticeable only for $Q>2m_t$, where the form factors $h_3^Z$ and $h_1^Z$ develop a relevant imaginary part, whereas for $Q$ values below $2m_t$, the new physics scenarios remain indistinguishable from SM predictions. Therefore, the contributions from imaginary parts of the form factors significantly impact the behavior of the different polarizations. 

\begin{figure}[!htb]
\begin{center}
\subfigure{}\includegraphics[width=6.25cm]{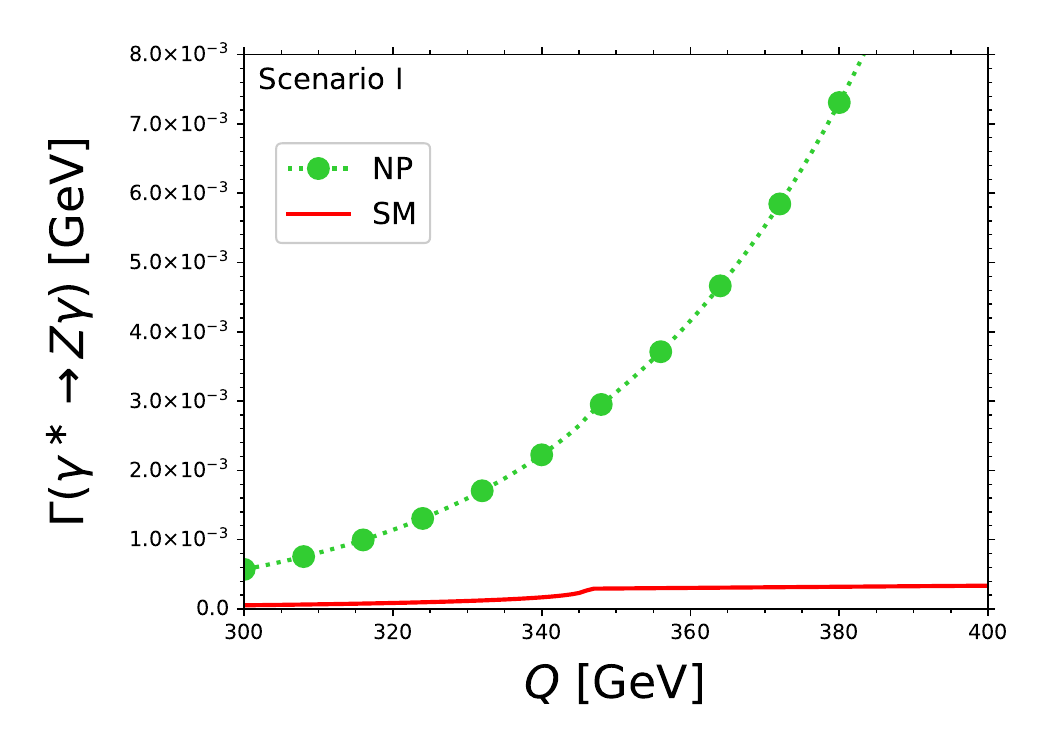}\hspace{-.73cm}
\subfigure{}\includegraphics[width=6.25cm]{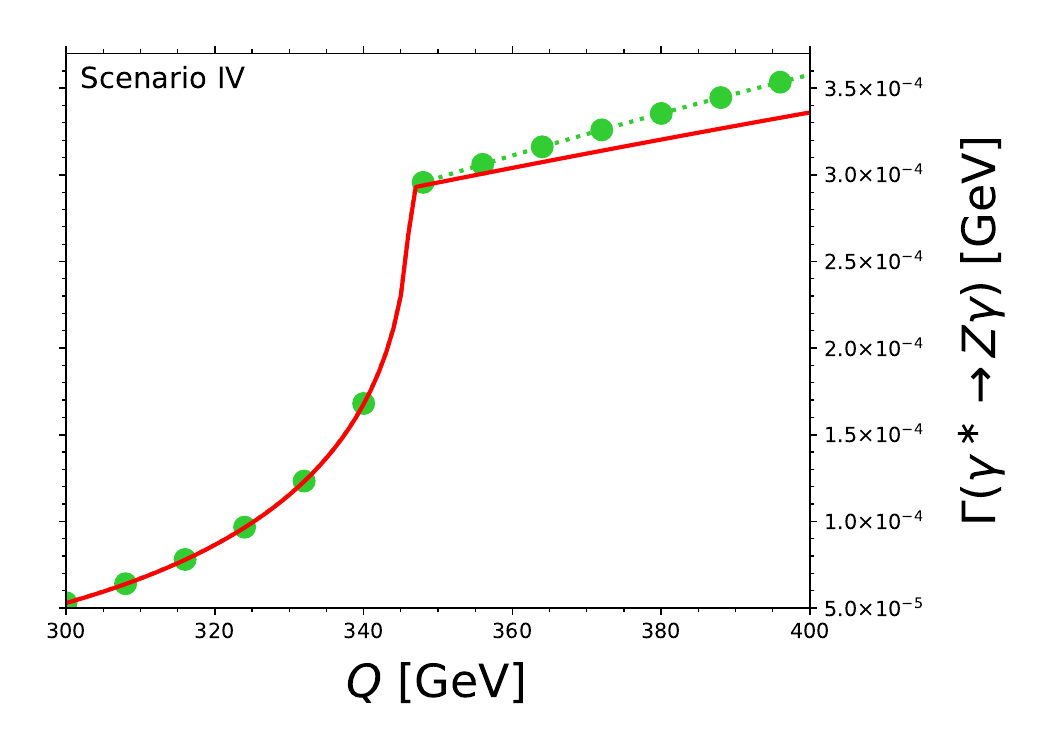}
\caption{We present the new physics (NP) and SM contributions to the partial decay width of the process $\gamma^\ast\rightarrow Z\gamma$.  \label{PlotsGamma11}}
\end{center}
\end{figure}

For completness, the new physics contributions (NP) decay widths for the process  $\gamma^\ast\rightarrow Z\gamma$ are shown in Fig. \ref{PlotsGamma11}. We focus on scenarios I and IV, as similar results are obtained for the other cases. Moreover, we include the one-loop SM contribution. The most significant values are reached in scenario I and of order $10^{-3}$. In scenario IV,  the contributions from the absorptive parts become notable for energies above $Q=2m_t$, allowing for a distinction between the new physics scenarios and the SM results. We observe that the $\gamma^\ast\rightarrow Z\gamma$ process is also sensitive to the imaginary parts of the form factors and shows significant deviations for new physics scenarios. Furthermore, the partial widths for this process are larger than those for an off-shell $Z$ boson.

Finally, for energies below 300 GeV, the behavior of the polarized partial widths is similar to that observed in Fig. \ref{PlotsZ1} and \ref{PlotsGamma11}. However, the corresponding values are about one order of magnitude lower. 

\subsection{Left-right asymmetry}

\begin{figure}[!htb]
\begin{center}
\includegraphics[width=14.2cm]{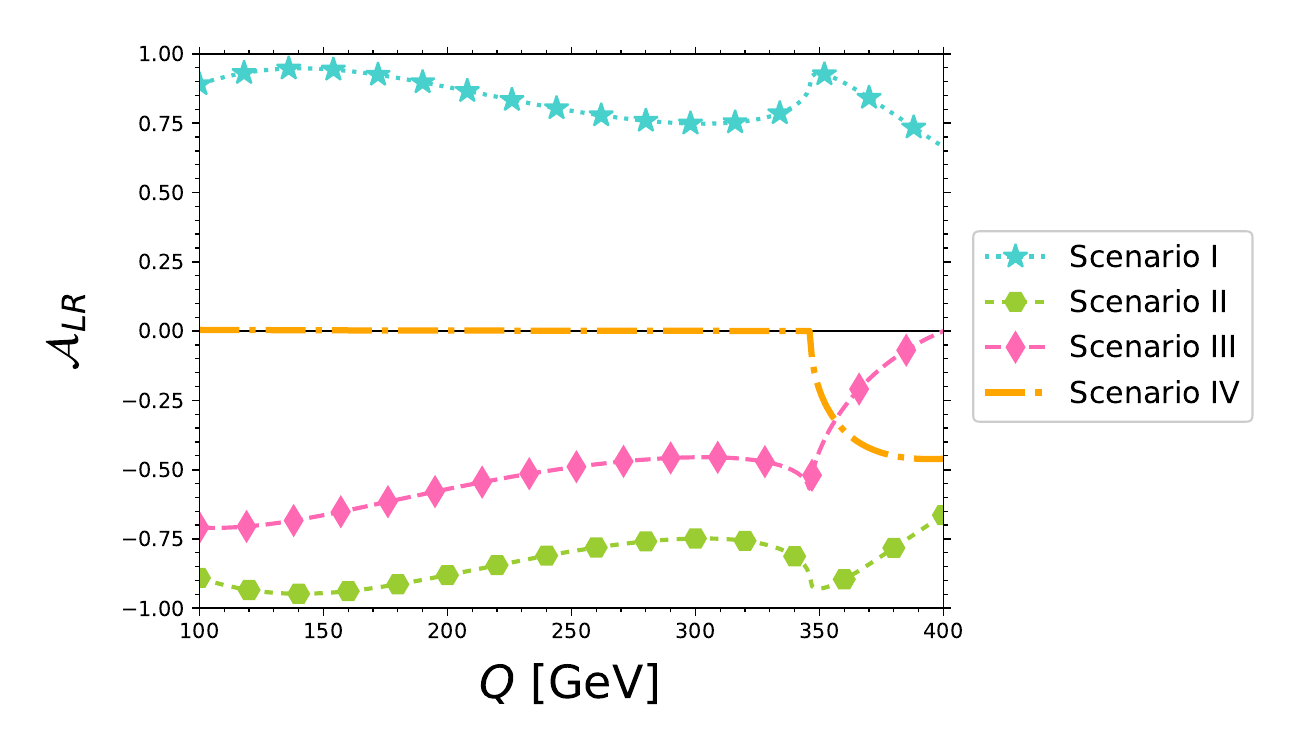}
\caption{$\mathcal{A}_{LR}$ asymmetry as function of $Q$ for the four new physics scenarios of Table. \ref{ta2}. The one-loop level SM contribution vanishes. \label{PlotsaAs1}}
\end{center}
\end{figure}

We now analyze the left-right asymmetries, which are sensitive to both the absorptive and $CP$-violating contributions of the TNGBCs. In Fig. \ref{PlotsaAs1}, we show the behavior of the $\mathcal{A}_{LR}$ asymmetry as a function of $Q$ considering the new physics scenarios in Table \ref{ta2}. In scenarios I and II, the real and absorptive parts of the $CP$-violating form factor $h_1^Z$ differ by a sign, whereas the SM contribution from $h_3^Z$ is negative in this energy region \cite{Gounaris:2000tb}. Hence, the terms in the numerator of Eq. \eqref{ALR11} combine constructively, resulting in an enhanced magnitude of the asymmetry. These values can approach the unity, indicating that the production rates of left-handed and right-handed  $Z\gamma$ pairs can be as significant as the total production of a transversely polarized $Z\gamma$ pair. In scenarios I-III, we note that the effects of the imaginary part of the $h_3^Z$ form factor become evident at energies above $2m_t$, as an abrupt change in the asymmetry behavior is observed. In scenario IV, for $Q<2m_t$,  the $h_3^{Z}$ and $h_1^Z$ remain real, and therefore the asymmetry vanishes. However, at higher energies, a sizeable absorptive part arises for both form factors, allowing $\mathcal{A}_{LR}$ to reach values around 0.5. Given that the asymmetry is identically zero within the SM, its detection would indicate new sources of $CP$ violation.

\section{Conclusions and outlook}\label{final}

In this work, we investigated the possibility of inducing new contributions to the $Z\gamma V$ ($V=Z$, $\gamma$) vertex through a generic model considering complex FCNC couplings mediated by the $Z$ boson. We find that new contributions to the $CP$-conserving form factor $h_3^Z$ are only induced in the $Z\gamma Z^\ast$ vertex, whereas no indication of $CP$ violation is observed in TNGBCs of the form  $Z\gamma V^\ast$ ($V=Z$, $\gamma$). Our result for $h_3^Z$ is presented in terms of the Passarino-Veltman scalar functions and can be reduced to the case of real FCNC couplings reported in the literature. To assess the magnitude of the new contributions to $h_3^Z$, we obtained constraints on the  $Z\overline{t}q$ couplings, using the current bounds on branching ratios $\mathcal{B}(t\rightarrow Zq)$ reported by the ATLAS collaboration. The resulting limits are:
$|g_{\text{V}}^{tu}|\text{,}|g_{\text{A}}^{tu}|\leqslant 0.007$ and  $|g_{\text{V}}^{tc}|\text{,}|g_{\text{A}}^{tc}|\leqslant 0.0095$.
The contributions to $h^Z_3$ can achieve values up order $10^{-7}$, three orders of magnitude smaller than those in the SM and MSSM. Although our results remain below the detection capabilities of the LHC, they could be sizeable at future colliders, such as the proposed 100 TeV $pp$ collider, which may reach a sensitivity to TNGBCs of order $10^{-7}$. 

The unpolarized and polarized partial decay widths of the $V^\ast\rightarrow Z\gamma$ process have also been calculated. This analysis incorporates complex form factors to investigate the effects of their imaginary parts and $CP$ violation. Our results indicate that the polarized observables are more sensitive to new physics effects due to their dependence on interference terms between the $CP$-conserving and $CP$-violating form factors, which cancel out in the unpolarized case. New physics scenarios can lead to considerable deviations in polarized partial decay widths compared to the SM predictions, especially when the imaginary parts of the form factors are relevant. Therefore, the absorptive parts notably affect the behavior of these observables. Of particular interest are the polarized partial widths $\Gamma^{\lambda}(Z^\ast\to Z\gamma)$  ($\lambda=L$, $R$), as they only arise through an off-shell $Z$ boson. Thus, examining transversely polarized $Z$ bosons presents a valuable opportunity to identify contributions from the $Z\gamma Z^\ast$ vertex. Such differentiation is not possible when observing unpolarized amplitudes. 

A new left-right asymmetry has been defined, which does not appear at the one-loop level in the SM. In specific new physics scenarios, its magnitude can be close to unity. Complex and CP-violating form factors are necessary to achieve a non-vanishing asymmetry. Finally, the measurement of $\mathcal{A}_{LR}$ would be evidence of new sources of $CP$ violation. 

\section*{Acknowledgments}
This work was supported by UNAM Posdoctoral Program (POSDOC). We also acknowledge support from  Sistema Nacional de Investigadores (Mexico). 

\section*{ORCID}

\noindent A.I. Hern\'andez-Ju\'arez- \url{https://orcid.org/0000-0003-0108-7358}

\noindent R. Gait\'an - \url{
https://orcid.org/0000-0001-7212-8722}

\noindent G. Tavares-Velasco - \url{https://orcid.org/0000-0002-0922-1163}

\appendix

\section{Photon polarization}\label{AppG}

By considering a polarized photon and a longitudinally polarized $Z$ boson, we can write $\Gamma^{0\lambda}(Z^\ast\to Z\gamma)$, where $\lambda$ denotes the polarization of the photon, as follows
\begin{align}\label{polZ0L}
\Gamma^{0L}(Z^\ast\rightarrow Z\gamma)=&\frac{e^2\left(Q^2-m_Z^2\right)^5}{256 \pi  Q^3 m_Z^{10}}
   \Bigg\{-2 Q^2 m_Z^2 \Big(\Big\{ {\rm Re}\big[h_2^Z\big]-{\rm Im}\big[h_4^Z\big]\Big\}
   \Big\{2
   {\rm Re}\big[h_1^Z\big]-2 {\rm Im}\big[h_3^Z\big]\nonumber\\
   &+{\rm Re}\big[h_2^Z\big]
   -{\rm Im}\big[h_4^Z\big]\Big\}+\Big\{ {\rm Re}\big[h_4^Z\big]+{\rm Im}\big[h_2^Z\big]\Big\} \Big\{2 {\rm Im}\big[h_1^Z\big]+2
   {\rm Re}\big[h_3^Z\big]+{\rm Re}\big[h_4^Z\big]+{\rm Im}\big[h_2^Z\big]\Big\}\Big)\nonumber\\
   &+m_Z^4 \Big(\Big\{2
   {\rm Re}\big[h_1^Z\big]-2
   {\rm Im}\big[h_3^Z\big]+{\rm Re}\big[h_2^Z\big]-{\rm Im}\big[h_4^Z\big]\Big\}{}^2\nonumber\\
   &+\Big\{2
   {\rm Re}\big[h_3^Z\big]+2
   {\rm Im}\big[h_1^Z\big]+{\rm Re}\big[h_4^Z\big]+{\rm Im}\big[h_2^Z\big]\Big\} {}^2\Big)\nonumber\\
   &
   +Q^4
   \Big(\Big\{{\rm Re}\big[h_2^Z\big]-{\rm Im}\big[h_4^Z\big]\Big\}{}^2+\Big\{ {\rm Re}\big[h_4^Z\big]+ {\rm Im}\left[h
   _2^Z\right]\Big\}{}^2\Big)\Bigg\},
\end{align}
whereas $\Gamma^{0R}(Z^\ast\rightarrow Z\gamma)$ is obtained from \eqref{polZ0L} after the replacement $h_{3,4}^Z\to -h_{3,4}^Z$.

For the $\gamma^\ast\rightarrow Z\gamma$ process, there are two distinct polarized partial decay widths, which can be written in terms of  $\Gamma^{0\lambda}(Z^\ast\rightarrow Z\gamma)$ as follows 
   \begin{align}\label{polGamma}
\Gamma^{0\lambda}(\gamma^\ast\rightarrow Z\gamma)=\frac{Q^4}{(Q^2-m_Z^2)^2}\Gamma^{0\lambda}(Z^\ast\rightarrow Z\gamma),\quad \lambda=L\text{, } R.
   \end{align}
   
   A new asymmetry involving only longitudinally polarized $Z$ bosons and can be defined as 
  \begin{equation}
\label{ALR2}
\mathcal{A}^{V}_{0LR}=\frac{\Gamma^{0L}(V^\ast\rightarrow Z\gamma)-\Gamma^{0R}(V^\ast\rightarrow Z\gamma)}{\Gamma^{0L}(V^\ast\rightarrow Z\gamma)+\Gamma^{0R}(V^\ast\rightarrow Z\gamma)}.
\end{equation}   
From Eqs. \eqref{polZ0L}-\eqref{polGamma}, we obtain
\begin{equation}\label{ALR22}
\mathcal{A}^{V}_{0LR}=\frac{g(Q)}{h(Q)},
\end{equation}
for either $V=Z$ or $V=\gamma$.
The $g(Q)$ and $h(Q)$ functions are given as
\begin{align}
g(Q)=&4 Q^2 m_Z^2 \Big\{ {\rm Im}\big[h_4^V\big]
   {\rm Re}\big[h_1^V\big]+\left({\rm Im}\big[h_3^V\big]+{\rm Im}\big[h_4^V\big]\right)
   {\rm Re}\big[h_2^V\big]-{\rm Im}\big[h_1^V\big]
   {\rm Re}\big[h_4^V\big]-{\rm Im}\big[h_2^V\big]
   \left({\rm Re}\big[h_3^V\big]+{\rm Re}\big[h_4^V\big]\right)\Big\}\nonumber\\
   &+2 m_Z^4
   \Big\{ \left(2 {\rm Im}\big[h_1^V\big]+{\rm Im}\big[h_2^V\big]\right) \left(2
   {\rm Re}\big[h_3^V\big]+{\rm Re}\big[h_4^V\big]\right)-\left(2
   {\rm Im}\big[h_3^V\big]+{\rm Im}\big[h_4^V\big]\right) \left(2
   {\rm Re}\big[h_1^V\big]+{\rm Re}\big[h_2^V\big]\right)\Big\}\nonumber\\
   &+2 Q^4
   \Big\{{\rm Im}\big[h_2^V\big] {\rm Re}\big[h_4^V\big]-{\rm Im}\big[h_4^V\big]
   {\rm Re}\big[h_2^V\big]\Big\}
\end{align}
and
\begin{align}
h(Q)=&-2 Q^2 m_Z^2 \Big\{ {\rm Im}\big[h_2^V\big]{}^2+2 {\rm Im}\big[h_1^V\big]
   {\rm Im}\big[h_2^V\big]+{\rm Im}\big[h_4^V\big]{}^2+2 {\rm Im}\big[h_3^V\big]
   {\rm Im}\big[h_4^V\big]+{\rm Re}\big[h_2^V\big]{}^2+{\rm Re}\big[h_4^V\big]{}^2\nonumber\\
   &+2
   {\rm Re}\big[h_1^V\big] {\rm Re}\big[h_2^V\big]+2 {\rm Re}\big[h_3^V\big]
   {\rm Re}\big[h_4^V\big]\Big\}+m_Z^4 \Big\{\left(2
   {\rm Im}\big[h_1^V\big]+{\rm Im}\big[h_2^V\big]\right){}^2+\left(2
   {\rm Im}\big[h_3^V\big]+{\rm Im}\big[h_4^V\big]\right){}^2\nonumber\\
   &+\left(2
   {\rm Re}\big[h_1^V\big]+{\rm Re}\big[h_2^V\big]\right){}^2+\left(2
   {\rm Re}\big[h_3^V\big]+{\rm Re}\big[h_4^V\big]\right){}^2\Big\}+Q^4
   \Big\{{\rm Im}\big[h_2^V\big]{}^2+{\rm Im}\big[h_4^V\big]{}^2+{\rm Re}\big[h_2^V\big]{}^
   2+{\rm Re}\big[h_4^V\big]{}^2\Big\}.
\end{align}
It is important to note that the numerator in Eq. \eqref{ALR22} only depends on the interference terms between $CP$-violating and $CP$-conserving form factors. Consequently, for a non-zero $\mathcal{A}^{V}_{0LR}$ asymmetry to exist, it is necessary to have complex and at least one non-zero $CP$-violating form factor.  Similar to the $\mathcal{A}^{V}_{LR}$ case, the $\mathcal{A}^{V}_{0LR}$ asymmetry vanishes in the SM.

\bibliographystyle{ws-ijmpa}
\bibliography{Biblio}

\end{document}